# Modelling politics in requirements engineering: adding emoji to existing notations

Rana Siadati, Paul Wernick and Vito Veneziano
School of Computer Science
University of Hertfordshire,
Hatfield, Hertfordshire
r.siadati@herts.ac.uk

*Abstract*—Notwithstanding several authors have recognised the conceptual key of "politics" as an important component in any Requirements Engineering (RE) process, practitioners still lack a pragmatic answer on how to deal with the political dimension: such an ability has become a mostly desirable but totally undetailed part of what we usually and vaguely refer to as "professional experience". Nor were practitioners given any suitable tool or method to easily detect, represent, control and if possible leverage politics. Authors argue that this issue could be successfully addressed and resolved if, when we map organisations against the system to be developed, we include power and politics in their "too human" and even emotional dimension.
A simple way to do so is to use emoji pictograms: most of them are part of a universal language, which requirements engineers could easily adopt and exploit to assess and produce models that include an extra layer of "political" information, without the need to actually introduce any new notation. A few examples of emoji-aware UML and organisational charts are hereby proposed, more as a platform to support communication and share reflections on how to deal with politics than as an actual technology to be adopted.

## I. Introduction

Over the last two decades, evolving concepts, practices and technologies in the software industry have greatly affected the requirements engineer's role. The "technical" component of such an evolution (supported by broader access to data and information, agile methods, new and more sophisticate tools and techniques, etc.), accompanied by other factors like time and memory loss, is not the only source of change in Requirements Engineering (RE) though, and the authors of this paper have argued elsewhere [23] that a quieter, deeper force seems more powerfully driving the underlying evolution of Requirements Engineering (RE) practice over time: political power.

Power and politics are far from being a new topic of reflection in Software RE [8], [3], [20], [19]: somehow inspired by political scientists (e.g. [7]) or management gurus (e.g. [14]), most definitions tend to converge into the idea the politics is the study of power at it happens and it mainly focuses on the "process of bargaining and negotiation that is used to overcome conflicts and differences of opinion" [19].

It is noticeable how non-technical aspects other than politics and power, such as commercial awareness and viability, finance and project management, have been better received by the practitioners as well as the academic community in software RE. We believe that this is either because they are considered to be more easily translatable into non-functional requirements, or just because it is nowadays accepted and expected that a requirements engineer should have both business knowledge and strong "soft skills" together with sound IT-related technical skills [11] in order to be able (and enabled) to negotiate with all stakeholders, even with the ones at the top of the pyramid (and negotiation is obviously part of the "politics" we are referring to).

However, we contend that over-simplified views and considerations of such aspects have become predominant in how we train requirements engineers: such views have likely contributed to a selective blindness for power dynamics and how they do not always propagate linearly, from top to bottom, but rather follow more complex patterns. For all the above reasons, it still may not seem to be "politically" too wise to use the word 'politics' in Software RE: politics has become such a difficult issue to handle (as it clashed against the traditional assumption that engineering is and should be, for its vocation and constitution, aseptic and neutral to the political dimension of human relationships) that practitioners were never given, nor they asked for, a sound pragmatic answer (or set of answers) on how to deal with politics alongside software development.

In last instance, it has been assumed by all parties that an ability to deal with (and sometimes to simply ignore) politics should be considered as a mostly desirable (but totally undetailed) part of what we usually refer to as "professional experience" [12]. This has been mirrored by the fact that the full set of (conceptual) tools available to requirements engineers (such modelling techniques, notations, pragmatic heuristics, protocols and guidelines, etc.) usually does not include anything which they could realistically adopt outside the speculative environment of academia for dealing with politics. So, how to break out of the impasse, given software development is deeply dominated by human factors: whilst we support that software construction processes can be, should be and usually are well-engineered, software developers and any other stakeholder in the development process are still creative human beings, who need to come to terms with unpredictable, sometimes turbulent, environments, made up of organisational and individual agendas, attitudes, personal interests, sympathies and conflicts, emotions, etc.

Because software engineering has not traditionally fully embraced the political (and indeed the human) dimension of its own nature, it is of little surprise that notations and tools in software engineering, as they are currently available to practitioners, are not suitable to represent politics in any useful (and usable) way. Also notations derived by other fields, like organizational charts, seem quite biased and unable to capture the actual political dynamics occurring within and under the official structure of any organisation. We argue that this issue could be successfully addressed and resolved by ensuring that, when we map organisations against the system we are expected to develop, we include power and politics in their "too human" and even emotional dimension.

## II. Politics and emotions in RE

Politics and power are very seldom an exercise in rationality [24] and even less in morality ([18], [1]): short-sighted self-interest, narcissism, sympathies, and the whole spectrum of human irrational (and sometimes unethical) behaviours are unfortunately more likely to affect politics than any rational analysis or ideal model. So, whilst in theory politics should be concerned with the fair pursuing of the "common good" of communities and organisations, and emphasis should therefore be placed on the "we all" perspective, in practice it is the "self" perspective, which individuals tend to adopt and give the higher priorities and consideration, creating a sort of epistemological conflict that resonates in any domain of human activity. Practitioners can anecdotally recall innumerable examples in which the "we all" is sacrificed on the altar of the "self", when they undertake the challenge to create software systems, which are supposedly expected to benefit organisations (of any sort), and eventually have to deal instead with individual members of that organisation who adopt the "self" (or only a subset of the "we") perspective.

However, whilst it is safe to assume it difficult (if not impossible) to know in detail -and likely of no utility to judge- whatever unobservable "political" motivations stakeholders are driven by alongside the requirements job, we argue that it is possible to focus on the observable political component of people behaviour, which carries a very visible component humans are usually unable to hid or control: emotions. Emotions (happiness, frustration, anger, fear, love, greed, etc.) can be revealed across many levels of human interactions, from the intentional words used and the consequential decisions made, to body language and up to chemical physiological reactions [15].

Whilst writers and poets can afford to describe emotions in detail, sometimes by using metaphors and/or accurate description of body and mind states, requirements engineers (need to) value speed and simplicity. Modelling is an attempt to translate accurate (but sometimes subjective) descriptions, which can be developed by using natural language, into simplified diagrammatic representations of the core abstraction of a problem, a system, a relationship. A simple way to model any captured emotions is by using emoji pictograms: although some of them are specific to a culture, the vast majority seem of them convey meaning rapidly across a shared universal language [2], by means of which requirements engineers operating world-wide could assess and quickly produce models that come with an extra layer of information, namely the political dimension, representing relationships and other relevant emotional information in an easy way, beyond any international translation difficulty and without the need to introduce any new notation.

## III. Including the emotional side of the political dimension in modelling

Modelling techniques based on pictures and diagrams, such as those included in UML (for example, use case diagrams) or others (like DFDs), tend to have one crucial advantage on words and human language, as philosophers of language and cognitive psychologists have well explained: they externalise the meaning they convey, differently from the written (or spoken) word, which -as a symbol- refers to a meaning we hold in our mind. As we have argued above, such techniques still come with an important flaw: they have been designed without considering the importance of capturing and analysing the political dimension of the context (or the broader system) they are to be engineered for. Hence, whilst analysts and designers focus on the structural component of the system to be developed, they proceed as if the system they are designing could be totally "abstracted" from reality and the political component it embeds: power relationships, sympathy, common interests, "clan belongingness" and other similar (and emotionally characterised) affecting factors, all of which we propose to simply denote as "political relationships". A few examples of emoji-aware UML are proposed, more as a platform to support communication and share reflections on how to deal with politics than as an actual technology to be adopted.

Example 1. A certain process (say representable as a level 2 DFD) is discussed between a manager (who is the "official" decision maker) and two workers: the requirements engineer has spotted that worker 1 is positively influencing the manager, and the other worker seems unhappy. However, the requirements engineer has been told that the second worker is going to be unhappy with the project and likely going to actively obstruct the work: the manager has de facto already decided not to be influenced (or bothered) by worker 2.

As a little insert in the DFD (say at the top left corner of the page), the situation could be represented as a quick sketch, as in Figure 1.

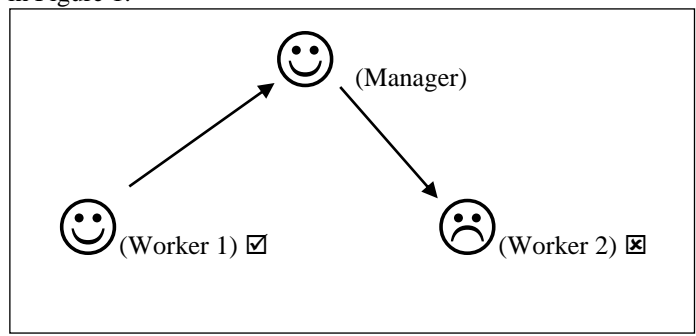

*Fig. 1: Political relationship (Scenario 1)*

This information is politically useful (and usable) for many reasons:

1. If the current requirements engineer leaves the project, any incoming substitute practitioner could immediately learn which sources are authoritative
2. Whilst obviously the manager is the "officially" decision maker, the political relationship says something about who the "underlying" decision maker is, and who actually influences whom
3. If the manager is unavailable for acting as -say- the product owner, practitioner knows whom s/he could talk to There is no need to represent an arrow pointing from the manager to worker 1: it is safe to assume that the manager does somehow exercise some power on worker 1. But it is more informative the arrow from worker 1 to manager. Such a simple extra notation is not immune from some sort of ambiguity: arrow from manager to worker 2 could be seen either as the ordinary influential relationship between the manager and the worker s/he coordinates, or it could be seen as an "active" power exercise "against" worker 2.

We argue that that both circumstances can be similarly dealt with by the same representation: whatever requirements worker 2 has an interest on, we better refer to the manager.

In other scenarios, there might be the need to consider the political dimension associated to the design of more atomic elements of a system: in such a case, political relationships can be highlighted (even partially, by simply representing one of the parties compounding the relationship) within the actual diagrammatic representation of that element, as per the following example.

Example 2. Methods of a certain class are discussed. One method x() is about reporting some complex information: a typical knowledge behaviour expensive to implement. Let's assume the above political relationship (as represented in Figure 1) stands, as this class affects the same manager and the same two workers.

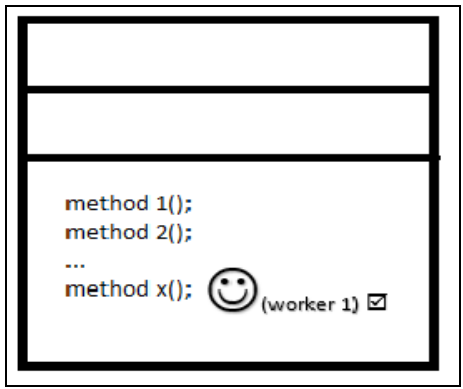

Fig. 2: Political characterisation within a UML Class (Scenario 2)

In this case, the political relationship is useful not just for allowing designers to immediately recognise whom to ask for more details about that method, but also which "political" priority to give, which readjustment has to be sought should any conflict arise between say implementation decisions, etc.

Two general points need to be taken into account: confidentiality of the political relationship and representation of more complex relationships than those provided here.
Confidentiality: no stakeholder should be made aware of the notation, as it could easily generate a butterfly effect in the organisation. We propose that political notation is only for the benefit of the requirements engineer.

Complexity: when a political relationship becomes more complex than, inevitably a more structured representation is needed: a typical example is when we want to represent stakeholders who are exercising their political influence behind (and sometimes even contradicting) the official organisational chart. In this case, little amendments to the simple representational tool can achieve the expected outcome, as for example in Figure 3.

It is important to underline that the political dimension does not need to be represented with the same extent of accuracy, unambiguity and systematic usage we usually demand from traditional engineering modelling techniques. The very same occasional presence of political notation could be seen as an important hint to the practitioner, who is then alerted and can pay special attention to individual requirements and/or stakeholder.

Also, practitioners need be aware that any description of the political dimension depends on how the personal interaction they might engage with stakeholders on a certain requirement, and therefore inherently carrying a subjective evaluation, which not necessarily is to be confirmed by another practitioner.

IV. CONCLUSION AND FUTURE WORK

Taking into account political relationships, as they occur between stakeholders and within complex organisations, can be of crucial importance for ensuring the success on any development project. At the moment, practitioners in RE can exercise their professional expertise by being "aware" of the political dimension, or by simply assuming the engineering process is politically neutral.

But, should they decide to adopt a more proactive approach to dealing with politics in RE, no simple notational technique is available to them. We are proposing a simple conceptual tools and an even simpler technique to capture, at least partially, important features of any political relationship within organisations and enrich their modelling outcomes. This could be of value for the individual practitioner, for whomever is due to substitute and/or replace the previous one, not just as a mean for identifying and describing potentially sources of problems and conflicts, but also for communication purposes.

The underlying idea of causality in political relationships (e.g., manager is affected by worker 1, but not from worker 2) seems naturally be referring to Bayes and similar approaches to formalise influence and dynamics of power. Whilst the formal representation of a Bayes-based political relationship is out of scope for the present study, we acknowledge this is an area worthy of further investigation and argue its outcomes could produce simple and yet effective tools, which practitioners can actually use in their daily activity.